\documentclass[reprint, superscriptaddress, aps, pra, nolongbibliography, amsmath, amssymb]{revtex4-2}

\usepackage[dvipdfmx]{graphicx}% Include figure files
\usepackage{dcolumn}% Align table columns on the decimal point
\usepackage{bm}% bold math
\usepackage[normalem]{ulem}
\usepackage{braket}
\usepackage{float}
\usepackage{amsmath,amssymb,bm}
\usepackage{ascmac}
\usepackage{array}
\usepackage{ulem}
\usepackage{hyperref}
\hypersetup{%
colorlinks=True,
linkcolor=blue, 
citecolor=blue,
urlcolor=blue
}
\begin{document}
\title{ZZ-Interaction-Free Single-Qubit-Gate Optimization in Superconducting Qubits}
\author{Shu Watanabe}
 \email{watanabe@qipe.t.u-tokyo.ac.jp}
 \affiliation{%
 Department of Applied Physics, Graduate School of Engineering, The University of Tokyo, 7-3-1 Hongo, Bunkyo-ku, Tokyo 113-8656, Japan
}%

\author{Yutaka Tabuchi}%
\affiliation{%
 RIKEN Center for Quantum Computing, Wako, Saitama 351-0198, Japan
}

\author{Kentaro Heya}%
\altaffiliation[Now at:]{
 IBM Quantum, IBM T.J. Watson Research Center, Yorktown Heights, New York 10598, USA
 }
\affiliation{%
 RIKEN Center for Quantum Computing, Wako, Saitama 351-0198, Japan
}%

\author{Shuhei Tamate}%
\affiliation{%
 RIKEN Center for Quantum Computing, Wako, Saitama 351-0198, Japan
}%

\author{Yasunobu Nakamura}%
\affiliation{%
 Department of Applied Physics, Graduate School of Engineering, The University of Tokyo, 7-3-1 Hongo, Bunkyo-ku, Tokyo 113-8656, Japan
}%
\affiliation{%
 RIKEN Center for Quantum Computing, Wako, Saitama 351-0198, Japan
}%

% \date{\today}% It is always \today, today,
%              %  but any date may be explicitly specified

\begin{abstract}
Overcoming the issue of qubit-frequency fluctuations is essential to realize stable and practical quantum computing with solid-state qubits. Static ZZ interaction, which causes a frequency shift of a qubit depending on the state of neighboring qubits, is one of the major obstacles to integrating fixed-frequency transmon qubits.  Here we propose and experimentally demonstrate ZZ-interaction-free single-qubit-gate operations on a superconducting transmon qubit by utilizing a semi-analytically optimized pulse based on a perturbative analysis. The gate is designed to be robust against slow qubit-frequency fluctuations. The robustness of the optimized gate spans a few MHz, which is sufficient for suppressing the adverse effects of the ZZ interaction. Our result paves the way for an efficient approach to overcoming the issue of ZZ interaction without any additional hardware overhead.
\end{abstract}

%\keywords{Suggested keywords}%Use show keys class option if keyword
                              %display desired
\maketitle

%\tableofcontents

% PRL does not have a section header
% \section{\label{sec:level1}First-level heading:\protect\\ The line break was forced \lowercase{via} \textbackslash\textbackslash}

% \section{Introduction}
\section{Introduction} 
Practical quantum information processing requires high-fidelity and stable quantum gate operations. In the platforms with solid-state qubits such as superconducting qubits and semiconductor spin qubits, a qubit frequency often drifts or fluctuates due to the influence of surrounding environments. The changes of qubit frequencies cause infidelity of their gate operations. 

Previous studies reported that two-level systems~(TLSs), charge noise, and parity switching are sources of the frequency fluctuations in superconducting qubits~\cite{Schlor2019, Christensen2019, Wilen2021, Tennant2022, Diamond2022}, which cause dephasing and gate errors due to the detuning. Most importantly, residual ZZ interaction between qubits, which changes the qubit frequency of interest depending on the neighboring qubit states, is an obstacle to integrating superconducting qubits. Multi-path couplings~\cite{Kandala2021, Zhao2021}, tunable couplers~\cite{Stehlik2021, Mundada2019, Xu2023}, weakly-driven couplers~\cite{Ni2022}, simultaneous ac Stark effects~\cite{Wei2022, Mitchell2021}, and utilizing low-degree graphs~\cite{Chamberland2020} have been proposed as countermeasures against the residual ZZ interaction.

To achieve high-fidelity quantum control in the presence of a device instability, such as a qubit frequency drift, a robust control scheme against the fluctuations is required. The concept of dynamically-corrected gates~\cite{Khodjasteh2009} is known for its ability to realize fluctuation-resilient quantum control. It was first introduced in nuclear magnetic resonance as a generalization of composite pulses~\cite{Tycko1983, Cummins2003, Cummins2000, Torosov2022}. We can regard the dynamically-corrected optimal gates as an extension of the spin echoes or dynamical decoupling~\cite{Tabuchi2017}, given that the control pulse ``refocuses'' faulty qubit dynamics due to fluctuating sources. Using a perturbative analysis with Magnus expansion~\cite{Haeberlen1968}, the dynamical-correction-based gate designs are free from sampling a large number of data sets of error strength unlike fully-numerical robust quantum-gate designs~\cite{Carvalho2021, Rembold2020, Le2023}. Incorporating the concept of dynamically-corrected gates into quantum optimal control theory~\cite{Werschnik2007, Khaneja2005, Rach2015, Reich2012} enables one to analytically~\cite{Dong2021, Dridi2020} or semi-analytically~\cite{Zeng2019, Buterakos2021, Gungordu2022, Deng2021, Hai2023} design robust quantum gates with smooth pulses. Beside dynamically-corrected gates, we also note that robust quantum-gate designs using deep reinforcement learning~\cite{Baum2021} have been proposed for solid-state qubits. It is also recognized that the echoed cross-resonance~\cite{Sheldon2016, Sundaresan2020} and the rotary pulses~\cite{Sundaresan2020, Jurcevic2021} are effective in mitigating the frequency shift due to the ZZ interaction during the execution of two-qubit cross-resonance gates between fixed-frequency transmon qubits. 

In this paper, we mitigate the negative impact of residual ZZ interaction between neighboring qubits on single-qubit gates by semi-analytically optimizing the pulse waveform based on dynamical correction. Our pulse optimization emphasizes the practicality of the pulse for use in experiments. By properly choosing the waveform ansatz, we avoid impractical pulse shapes with a large derivative. Another feature of our method lies in its universality. The optimal waveform is determined independently of the individual qubit parameters, and the experimental protocol for gate calibration is as simple as those for conventional waveforms such as Gaussian. This approach for ZZ-interaction suppression is distinct from other existing methods mentioned above since it requires no hardware overheads, such as additional circuit elements, flux bias lines, or drive channels.

\section{DYNAMICALLY-CORRECTED GATES}
\label{app:dcg}
 Our strategy for optimizing a single-qubit gate robust against detuning is based on the concept of dynamically-corrected gates~(DCG)~\cite{Khodjasteh2009, Zeng2019, Buterakos2021, Gungordu2022, Deng2021, Hai2023}. In the DCG-based synthesis, we begin by classifying the Hamiltonian of the system into an unperturbed term $\hat{H}_{0}(t)$ and an error term $\hat{H}_{\mathrm{error}}(t)$ as
\begin{align}
    \hat{H}_{\mathrm{total}}(t) = \hat{H}_{0}(t) + \hat{H}_{\mathrm{error}}(t).
\end{align}
With the classification, we separate the overall time propagator $\hat{U}_{\mathrm{total}}(t)$ into non-perturbative dynamics $\hat{U}_{\mathrm{0}}(t)$ and faulty dynamics $\hat{U}_{\mathrm{I}}(t)$ as 
\begin{math}
    \hat{U}_{\mathrm{total}}(t) = \hat{U}_{\mathrm{0}}(t)\hat{U}_{\mathrm{I}}(t)
\end{math}
where we define
\begin{align}
    \dot{\hat{U}}_{\mathrm{0}}(t) = -i\hat{H}_{\mathrm{0}}(t)\hat{U}_{\mathrm{0}}(t)
\end{align}
and
\begin{align}
    \dot{\hat{U}}_{\mathrm{I}}(t) = -i\hat{H}_{\mathrm{I}}(t)\hat{U}_{\mathrm{I}}(t)
\end{align}
by using the toggling-frame Hamiltonian 
\begin{align}
    \hat{H}_{\mathrm{I}}(t) = \hat{U}_{\mathrm{0}}^{\dagger}(t)\hat{H}_{\mathrm{error}}(t)\hat{U}_{\mathrm{0}}(t).
\end{align}
It is useful to approximate the faulty dynamics over a time interval $T$ by defining the effective time-independent ``average'' Hamiltonian:
\begin{align}
    \hat{U}_{\mathrm{I}}(t) \simeq \exp(-i\hat{\overline{H}}T).
\end{align}
The average Hamiltonian can be expressed by Magnus expansion~\cite{Haeberlen1968, Tabuchi2017}
\begin{align}
    \hat{\overline{H}} = \sum_{n}\hat{\overline{H}}^{(n)}.
\end{align} 
The lowest and second-lowest orders of Magnus expansion are given by
\begin{align}
    \hat{\overline{H}}^{(1)} &= (1/T)\int_0^Tdt\,\hat{H}_{\mathrm{I}}(t),\\
    \hat{\overline{H}}^{(2)} &= \frac{-i}{2T}\int_0^T\int_0^tdt\,dt'\left[\hat{H}_{\mathrm{I}}(t), \hat{H}_{\mathrm{I}}(t')\right].
\end{align}
The precision of the approximation is guaranteed when $\|\hat{H}_{\mathrm{error}}T\|\ll1$. By designing the unperturbed term $\hat{H}_{\mathrm{0}}(t)$ so that the average Hamiltonian $\hat{\overline{H}}$ vanishes, we can consider that the effect of the error term on the system's time evolution is stroboscopically negligible at time $T$. Since the Magnus expansion is defined with the integrals, whether the average Hamiltonian is zero or not is independent of the norm of the error terms and depends only on the functional form of $\hat{H}_{\mathrm{error}}(t)$ and $\hat{H}_{\mathrm{0}}(t)$. 

Under a single-qubit control drive at the qubit frequency and with the rotating-wave approximation, the unperturbed term $\hat{H}_{\mathrm{0}}(t)$ becomes
\begin{align}
    \hat{H}_{\mathrm{0}}(t) = \Omega_x(t)\frac{\hat{\sigma}_x}{2} + \Omega_y(t)\frac{\hat{\sigma}_y}{2}.
    \label{eq:drive}
\end{align}
Therefore, a dynamically-corrected single-qubit gate can be designed in two steps: first, we decide which errors to gain robustness against, and second, we optimize the waveforms $\Omega_x(t)$ and $\Omega_y(t)$ so that the average Hamiltonian in the toggling frame vanishes. As discussed above, the vanishment of the average Hamiltonian depends only on the functional form of the error term and the drive waveform, not on the norm of the error terms. In other words, by using the toggling frame and average Hamiltonian, we can naturally acquire robustness against the amplitude of the error. Therefore, our strategy is to find through a numerical optimization the optimal control waveforms $\Omega_x(t)$ and $\Omega_y(t)$ that simultaneously satisfy two conditions
\begin{equation}
 \hat{U}_{0}(T) = \hat{U}_{\mathrm{target}}
 \label{eq:condition1}
 \end{equation}
 and 
 \begin{equation}
 \hat{\overline{H}} = 0,
 \label{eq:condition2}
\end{equation} 
where $\hat{U}_{\mathrm{target}}$ is the unitary that expresses the target gate. The former condition aims to achieve the desired gate operation, and the latter aims to acquire robustness against errors.

\section{Pulse synthesis}
We synthesize an optimal waveform of a single-qubit $X_{{\pi}/{2}}$ gate robust against the time-independent or slowly-varying detuning of the qubit frequency modeled as 
\begin{math}
    \hat{H}_{\mathrm{error}} = \xi\hat{\sigma}_z/2.
\end{math}
Because of the non-commutativity of $\hat{\sigma}_z$ and $\hat{\sigma}_x$, the in-phase component of the drive $\Omega_x(t)$ is sufficient for mitigating the detuning error that we focus on. This simplification reduces the number of optimization parameters and accelerates the convergence. By using the unitary operator
\begin{math}
    \hat{U}_0(t) = \mathcal{T}\exp \left(-i\int_0^t\Omega_x(\tau)\mathrm{d}\tau\frac{\hat{\sigma}_x}{2}\right),
\end{math}
where $\mathcal{T}$ stands for the time-ordered product, the error Hamiltonian on the toggling frame
\begin{math}
\hat{H}_{\mathrm{I}}(t) = \hat{U}_{\mathrm{0}}^{\dagger}(t)\hat{H}_{\mathrm{error}}\hat{U}_{\mathrm{0}}(t)
\end{math}
is derived as follows:
\begin{align}
    \hat{H}_{\mathrm{I}}(t) = \xi\left(\cos \Theta(t)\,\frac{\hat{\sigma}_{z}}{2} + \sin \Theta(t)\,\frac{\hat{\sigma}_{y}}{2}\right)
\end{align}
where $\Theta(t) = \int_0^t\Omega_x(\tau)\mathrm{d}\tau$. We then approximate the faulty dynamics introduced by the error term using Magnus expansion based on the average Hamiltonian theory~\cite{Ernst1990}. The exact expression of Magnus expansion is found in Appendix~\ref{app:dcg}. The deviation of the dynamics from the ideal one can be neglected when the drive is designed such that the error Hamiltonian vanishes stroboscopically under the truncation in the Magnus expansion. We numerically optimize the drive waveforms $\Omega_x(t)$ by adopting a waveform ansatz expressed as the product of polynomials and cosine:
\begin{align}
    \Omega_{x}(t) = \sum_{n}a_n^{(x)}\left(t-\frac{T}{2}\right)^n\cdot\cos^2\left(\frac{\pi}{T}\left(t-\frac{T}{2}\right)\right),
    \label{eq:ansatz}
\end{align}
where $\left\{a_n^{(x)}\right\}$ are the parameters to be optimized and $T$ is the gate length. We suppose that the pulse of Eq.~(2) is applied during the period $0 \le t \le T$, and otherwise $\Omega_x(t) = 0$.
 This ansatz forces the pulse and its derivative to start and end at zero. Considering the use of the DRAG correction, as discussed later, this is essential to enhance the practicality of the optimized gate in experiments since we can minimize the probability of non-adiabatic transition. The careful choice of the waveform ansatz enables us to realize relatively fast and high-fidelity single-qubit gates compared to the previous work~\cite{Hai2023}. The pulse optimization is executed to satisfy two conditions: (i) the overall time evolution of the qubit is identical to the target gate, and (ii) the error Hamiltonian on the toggling frame is averaged to zero. 

 The optimization problem is to minimize a cost function 
\begin{math}
    C = 1-\frac{1}{2}\left|\mathrm{Tr}\left(\hat{U}_{\mathrm{target}}^{\dagger}\hat{U}_{\mathrm{0}}(T)\right)\right|
\end{math}
subject to  
\begin{math}
 \hat{\overline{H}} = 0.
\end{math} 
We treat this constrained optimization problem as a multi-objective optimization problem to minimize 
\begin{multline}
    C_{\mathrm{all}}\left(\left\{a_n^{(x)}\right\}\right) = w\cdot {C}_{\mathrm{fidelity}}\left(\left\{a_n^{(x)}\right\}\right) \\
    \hspace{1cm}+ (1-w)\cdot  C_{\mathrm{robust}}\left(\left\{a_n^{(x)}\right\}\right),
    \label{eq:costfunction}
\end{multline}
where the parameters $\left\{a_n^{(x)}\right\}$ are defined as Eq.~(2) in the main text to parametrize the shape of the $X$-control pulse. We define two parts of the cost function as
\begin{align}
    C_{\mathrm{fidelity}}\left(\left\{a_n^{(x)}\right\}\right) = \left|\Theta(T) - \frac{\pi}{2}\right| 
\end{align}
and
\begin{multline}
    C_{\mathrm{robust}}\left(\left\{a_n^{(x)}\right\}\right) \\
    = \frac{1}{T}\left(\left|\int_0^T\cos \Theta(t)\mathrm{d}t\right| + \left|\int_0^T\sin \Theta(t)\mathrm{d}t\right|\right),
    \label{eq:cost2}
\end{multline}
where $\Theta(t) = \int_0^t\Omega_x(\tau)\mathrm{d}\tau$ is used for simplification. Note that Eq.~\eqref{eq:cost2} characterizes the $L_1$ norm of the lowest-order average Hamiltonian calculated from Eq.~(1) in the main text. These two parts of the cost function correspond to the conditions Eqs.~\eqref{eq:condition1} and \eqref{eq:condition2} explained above, respectively. The weight factor $w$ in Eq.~\eqref{eq:costfunction}, which determines the balance between maximizing fidelity and minimizing the norm of the average Hamiltonian, is deliberately chosen between 0 and 1 to achieve the desired optimization. The value of $w$ is set to be 0.999 in our optimization since we empirically find that it is more likely to make both terms in the cost function Eq.~\eqref{eq:costfunction} smaller when w is close to 1.

The parameters of the synthesized waveform expressed with the ansatz Eq.~(2) in the main text are listed in Table~\ref{table:parameter}. Using higher-degree polynomials increases the freedom of expression of the waveform but likely results in a larger derivative of the waveform, which may increase the probability of non-adiabatic transitions. In addition, instruments impose bandwidth limitations in experiments. We use polynomials up to the second order~($n\le 2$). We also fix the first-order coefficient~($n = 1$) to zero to synthesize a time-symmetric waveform, which is not mathematically essential. The gate duration $T$ is chosen to be 40~ns. The initial guess of the parameters are $a_0^{(x)} = \pi/T$ and $a_n^{(x)} = 0$ for $n\ne 0$. We utilize the L-BFGS method~\cite{Nocedal1980} for the optimization. The numerical integrals necessary for the cost evaluation are performed by dividing 1~ns into 64 equal fractions.
\begin{table}[htbp]
 \caption{Parameters of the optimal $X_{{\pi}/{2}}$ pulse robust against detuning.}
 \label{table:parameter}
 \centering
  \begin{tabular}{lWc{1.5cm}}
   \hline \hline
   $n$ & $a_n^{(x)}$\\
   \hline 
   0 & $0.31831$ \\
   2 & $-0.00515$  \\
   \hline \hline
  \end{tabular}
\end{table}

It is worth mentioning that since the cost function Eq.~\eqref{eq:costfunction} does not include any information about the system, such as the qubit frequency or magnitude of the frequency drift, the obtained values in Table~\ref{table:parameter} are universal and commonly applicable to any qubit that can be expressed by this model as long as the convergence condition for the Magnus expansion, $\|\xi T\|\ll1$, is fulfilled. 

\section{Experiment} 
We use fixed-frequency transmon qubits Q0 and Q1 in the experiments. For detailed information on the whole device and the qubits used in the experiments, see Appendix~\ref{app:exp}.

\begin{figure}[htbp]
  \centering
  \includegraphics[width=8.6cm,pagebox=cropbox, clip]{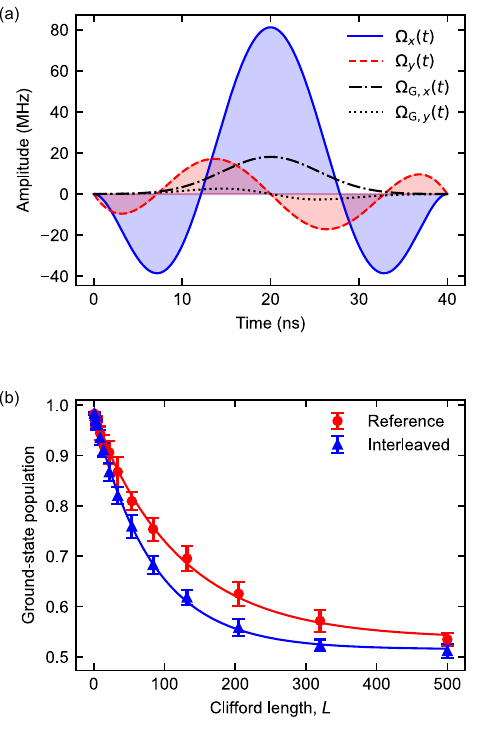}
  \caption{(a) Optimized waveform $\Omega_x (t)$~(solid) for the $X_{{\pi}/{2}}$ gate with calibrated DRAG correction $\Omega_y (t)$~(dashed). The real~($x$) and imaginary~($y$) parts of the Gaussian waveform $\Omega_{G,x} (t)$ with calibrated DRAG correction $\Omega_{G,y} (t)$ for the same gate are shown as dash-dotted and dotted lines, respectively. (b) Interleaved randomized benchmarking~(IRB) of the $X_{{\pi}/{2}}$ gate using the optimal waveform at zero detuning. The plot is averaged over 40 random Clifford sequences per point. The measured error rate per gate~(EPG) is $1.9(2)\times 10^{-3}$. }
  \label{fig:result}
\end{figure}
 First, we use the target qubit Q0 to evaluate the gate fidelity and the robustness against detuning error. We set the pulse length $T$ to be 40~ns and construct the optimized pulse with the polynomials up to $n=2$. We also apply the Derivative Removal by Adiabatic Gate~(DRAG) with $Y$-only correction, mentioned in Ref.~\citenum{Gambetta2011} since we empirically find through numerical simulations that the impact of the use of $Y$-only DRAG on the robustness of the optimized gate against the detuning is small. We emphasize that the optimized waveform is universal in the sense that it is independent of the individual qubit parameters; we only need to experimentally calibrate the absolute pulse amplitude and DRAG weight for each qubit. Thus, the calibration protocol is as simple as the one for conventional waveforms like Gaussian. It should also be noted that the maximum amplitude of the optimal waveform is about four times larger than that of the Gaussian waveform for the identical gate length, as shown in Fig.~\ref{fig:result}(a). In general, the more degrees we include in the polynomials in the ansatz, the more robustness we gain. However, the time derivative of the waveform becomes larger and non-adiabatic transitions become more significant. This trade-off is a factor that needs to be considered when one seeks to realize fast and robust quantum gates. An analysis of leakage error to the higher excited state is shown in Sec.~\ref{sec:leak}.
 
\begin{figure}[htbp]
  \centering
  \includegraphics[width=8.6cm,pagebox=cropbox, clip]{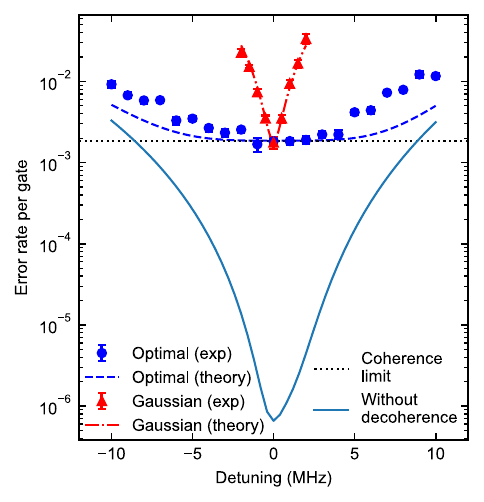}
  \caption{EPG extracted from IRB versus drive detuning for the $X_{{\pi}/{2}}$ gate with the optimized~(blue circle) and Gaussian~(red triangle)  waveforms with the theoretical predictions~(dashed line for the optimal and dashed-dotted line for the Gaussian waveforms). The solid line represents the theoretical limit to the gate error with infinite coherence times. The dotted line represents the theoretical coherence limit to the gate error.}
  \label{fig:detuning}
\end{figure}

 Next, we evaluate the gate fidelity and robustness of the optimized pulse based on randomized benchmarking~(RB)~\cite{Magesan2011, Magesan2012}. See Appendix~\ref{app:cal} for the calibration details. Figures~\ref{fig:result}(a) and~(b) are the calibrated waveform and the result of interleaved randomized benchmarking~(IRB), respectively. In the RB sequence, all single-qubit Clifford gates consist of two $X_{{\pi}/{2}}$ gates and three virtual-Z gates~\cite{McKay2017}. 
The gate fidelity measured by the IRB is $99.81\pm 0.02\%$. We note that this fidelity is mainly limited by the coherence of the qubit. The energy relaxation time $T_1$ and the echo dephasing time $T_2$ of Q0 are 13~$\mu$s and 10~$\mu$s, respectively. The theoretical coherence limit of the gate fidelity is numerically obtained using measured $T_1$ and $T_2$. We also note that the numerically-obtained error rate per gate of the optimized waveform without decoherence is less than $10^{-6}$, which implies that there is significant room for achieving much better fidelity by improving the coherence time. In Fig.~\ref{fig:detuning}, we show the error rate per gate~(EPG) versus drive detuning to investigate the robustness of the optimized waveform against the detuning. As our qubit is a fixed-frequency qubit, we detune the frequency of the driving pulse instead of the qubit frequency. Here we apply the intentional detuning only to the interleaved gates. The robustness is compared with the widely-used DRAG-optimized Gaussian pulse with the same pulse length of 40~ns, and shows a significant improvement. The increase in EPG is well suppressed even with a few-MHz detuning. The range of robustness is consistent with the condition $\lvert\xi T\rvert\ll 1$, which justifies the lowest-order Magnus expansion.

Considering that the designed gate maintains high fidelity over the wide bandwidth and that the typical magnitude of frequency drift due to TLSs and parity switching in superconducting transmon qubits is at most an order of MHz~\cite{Schlor2019, Wilen2021, Tennant2022, Diamond2022}, such broadband quantum gates can be an effective countermeasure to the degradation of control fidelity due to these slow fluctuations. Although transmon qubits are generally insensitive to the charge noise, it cannot be ignored when one implements quantum gates in non-computational spaces such as $\ket{\mathrm{e}}$--$\ket{\mathrm{f}}$. Such control has recently been found useful for an the efficient decomposition of Toffoli gates on qutrits~\cite{Nikolaeva2022, Luo2023}. Aside from superconducting qubits, another platform that can be benefited from this optimized gate robust against detuning is semiconductor spin qubits. In silicon-based spin qubits, it is reported that the magnitude of slow drift in the qubit frequency is on the order of hundreds of kHz~\cite{Takeda2022, Noiri2022}. In addition, it is commonly known that slow dynamics of nuclear spin bath affects the energy splitting of an electron spin via hyperfine interaction. Maintaining control fidelity under such frequency fluctuations is a prerequisite for in quantum computing.

Next, we study the impact of the residual ZZ interaction on the synthesized optimal single-qubit gate, using a spectator qubit Q1 which is capacitively coupled to the target qubit Q0, as shown in the Appendix~\ref{app:exp}. Under the ZZ interaction with a spectator qubit, $\hat{H}_{\mathrm{error}}$ is expressed as 
\begin{equation}
\hat{H}_{\mathrm{error}} = \xi_{ZZ}\frac{\hat{{\sigma}}_{{z0}}\otimes\hat{\sigma}_{{z1}}}{2}.
\end{equation}
Here $\hat{{\sigma}}_{{z0}}$ and $\hat{\sigma}_{{z1}}$ are the Pauli-Z operator of the target and the spectator qubits, respectively, and $\xi_{ZZ}$ denotes the ZZ coupling strength in between.
Assuming that the state of the adjacent qubit is unchanged during the gate execution, we can reduce $\hat{H}_{\mathrm{error}}$ to a frequency shift of the target qubit:
\begin{equation}
\hat{H}_{\mathrm{error}} = \chi\frac{\hat{\sigma}_{z0}}{2},
\end{equation}
 where $\chi = \pm\,\xi_{ZZ}$. Note that this expression scales when the number of adjacent qubits increases if we assume that the states of all the adjacent qubits are unchanged.
 In the present device, the ZZ interaction $\xi_{\mathrm{ZZ}}$ between Q0 and Q1 is $2\pi\times 0.73$~MHz.  Since our single-qubit gate length $T$ is 40~ns, satisfying $|\chi T|\ll1$, the legitimacy of the approximation with the lowest-order Magnus expansion is guaranteed.

\begin{figure}[H]
  \centering
  \includegraphics[width=8.6cm,pagebox=cropbox, clip]{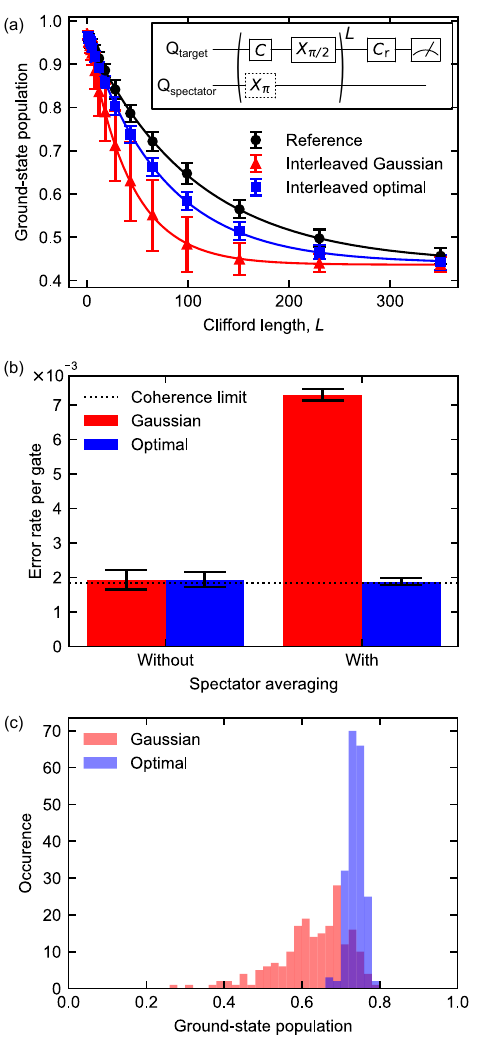}
  \caption{(a) Interleaved randomized benchmarking with spectator averaging for Gaussian and optimal waveform. The plot is averaged over 200 random Clifford sequences and 2000 shots per point. The gate schematic is shown in the inset. (b) Comparison of the error rates per gate with and without averaging the state of the spectator qubit. The dotted line indicates the coherence limit to the EPG. (c) Distributions of ground-state population for the Gaussian and optimal waveforms for Clifford length $L=43$ in the spectator-averaging RB.}
  \label{fig:rb_sr}
\end{figure}

We incorporate an RB-like methodology to quantify the impact of the spectator error. In the sequence, we run the standard IRB on the target qubit. In addition, we randomly apply single-qubit $X_{\pi}$ pulses to the spectator qubit in synchronization with the reference Clifford gate to the target qubit. By averaging the result over a sufficiently large number of samplings, the spectator qubit's expectation value right before the $X_{\pi/2}$ pulse on the target qubit is averaged to $\overline{\left<\hat{\sigma}_z\right>}=0$. The deviation of the fidelity measured with this averaging from the one measured without operation to the spectator qubit shows the influence of the changes of the spectator's state, that is, the impact of the residual ZZ coupling. In the following, this technique is referred to as spectator-averaging RB. The gate schematic is shown in the inset of Fig.~\ref{fig:rb_sr}(a).

Figure~\ref{fig:rb_sr}(a) shows the results of spectator-averaging RB for the optimal and Gaussian waveforms. The comparison of error rates with and without the spectator-averaging shows that the averaging significantly affects the fidelity of the Gaussian $X_{{\pi}/{2}}$ gate, as shown in Fig.~\ref{fig:rb_sr}(b). The EPG with spectator-averaging is about four times worse than that without averaging. On the other hand, when the optimal waveform is employed, the EPG with spectator-averaging is $1.9(1)\times 10^{-3}$, which is almost equal to the EPG obtained with the standard IRB. Figure~\ref{fig:rb_sr}(c) shows the distributions of the ground-state population after applying 43 Clifford gates in the spectator-averaging RB for the Gaussian and optimal waveforms, respectively. We observe a significantly larger spread in the distribution for the Gaussian waveform than the optimal one. The asymmetric shape of the distribution for the Gaussian gate is a consequence of the systematic error~\cite{Ball2016, Ganzhorn2020}, which in this case is the spectator-dependent phase error due to the ZZ interaction. On the other hand, the relatively small spread in the distribution for the optimized gate indicates that it is insensitive to such phase error. Therefore, the optimized gate robust to detuning can reduce the negative impact of the ZZ interaction between the target and spectator qubits. As the magnitude of the ZZ interaction between qubits in typical transmon integrated circuits is several tens or hundreds of kHz~\cite{Ni2022, Wei2022} and the optimized gate has broadband robustness over a few MHz of detuning, the scheme should work even when the number of adjacent qubits increases and the target qubit is exposed to ZZ interactions from them. The robustness contributes to the architectures with high qubit connectivity, including square lattices, while other issues, such as frequency collisions and two-qubit gates, remain. Since this implementation of a robust quantum gate does not require any additional wiring, circuit element, or drive channel, it serves as an alternative approach to tackling the issue of ZZ interaction.

Furthermore, the single-qubit gate robust to ZZ-interaction is potentially useful for quantum computation scheme controlling always-on internal interactions through decoupling and selective recoupling~\cite{Tsunoda2020, Leung2000, Jones1999, Parra-Rodriguez2020}. $X$ gates are necessary for decoupling and recoupling sequences, which are not trivial under always-on coupling. Our optimized gate robust to ZZ-interaction can be applied to implement high-fidelity $X$ gates in such architectures.

\section{LEAKAGE ANALYSIS}
\label{sec:leak}
\begin{figure}[htbp]
  \centering
  \includegraphics[width=8.6cm,pagebox=cropbox, clip]{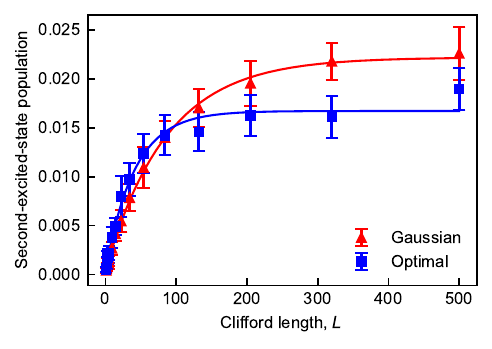}
  \caption{Leakage randomized benchmarking~(LRB) of the $X_{\pi/2}$ gate using the Gaussian and optimal waveforms. The plot is averaged over 40 random Clifford sequences per point. The measured leakage rate per gate is $1.3(1)\times10^{-4}$ and $2.0(2)\times10^{-4}$ for the Gaussian and optimal waveforms, respectively. The solid lines represent the fits to the decay model.}
  \label{fig:lrb}
\end{figure}

We evaluate the leakage error to the second excited state by leakage randomized benchmarking~(LRB)~\cite{Wood2018}. The pulse sequence of LRB is the same as standard RB. We fit the population of the second excited state averaged over 40 random Clifford sequences as a function of the sequence length $L$ to the decay model $p_2(L) = Ap^L+B$. The leakage rate per gate~(LPG) is then computed as 
\begin{align}
    l = B(1-p)/2
\end{align}
since all single-qubit Clifford gates can be represented by two $X_{\pi/2}$ gates and three virtual-Z gates. In Fig.~\ref{fig:lrb}, we show LRB curves for the Gaussian and optimal waveforms. The LPGs measured by LRB are $1.3(1)\times10^{-4}$ and $2.0(2)\times10^{-4}$ for the Gaussian and optimal waveforms, respectively. On the other hand, numerically-obtained LPGs are $1.3\times10^{-4}$ and $1.5\times10^{-4}$ for the Gaussian and optimal waveforms, respectively. The optimal waveform induces slightly more leakage to the non-computational state than the simulation. We suppose that this is because of the influence of surrounding qubits that is not included in the simulation. We also obtain the lower bound on the LPG due to finite-temperature heating~\cite{Zijun2016} by numerical simulation. Using measured $T_1$, $T_2$ and the thermal excitation rate of the qubit, we calculated the lower bound as $1.3\times10^{-4}$. Furthermore, we simulate the LPG of the optimal waveform without decoherence. The obtained LPG without decoherence is less than $10^{-6}$, implying that the leakage error would not limit the gate fidelity until the coherence time becomes by far longer. Therefore, although the experiment shows that the optimal waveform induces slightly more leakage to the non-computational state than the Gaussian waveform of the same length, the leakage error is basically dominated by incoherent state transfer due to finite temperature and not limiting the gate fidelity.

\section{conclusions}
In this paper, we propose and experimentally demonstrate suppression of the adverse effect of ZZ interaction on single-qubit gates in transmon qubits using semi-analytically optimized waveform. We optimize the pulse waveforms to obtain robustness by utilizing dynamically-corrected gates. Our pulse optimization emphasizes the practicality of the pulse for use in experiments. By properly choosing the waveform ansatz, we avoid impractical pulse shapes with a large derivative or discontinuity. Our approach is distinct from other existing methods for mitigating the ZZ interaction, such as tunable or multi-path couplers and simultaneous ac Stark effects since it requires no additional circuit elements, wiring, or drive channels.

The error rate per gate of our optimized single-qubit gate is $1.9(2)\times 10^{-3}$, which is mainly limited by decoherence of the qubit. The error rate per gate keeps unchanged even under a few MHz of detuning. 

The broadband robustness of the optimized gate is exploited to mitigate ZZ interaction during single-qubit gate operation. Using a qubit pair with the ZZ interaction of 0.73~MHz, we demonstrate significant improvement in the gate fidelity in comparison to the Gaussian waveform. 

Our results pave the way for a hardware-efficient ZZ-interaction-free quantum control in superconducting qubits. It also serves as a novel countermeasure against the issue of frequency fluctuations in quantum computation based on solid-state qubits.

% Finally, since the optimal echoes provide a highly scalable and general way to design robust quantum gates against various kinds of systematic error and are fully compatible with other quantum systems, it paves the way for a new solution to deal with the problems such as frequency collisions and device instability in a wide range of quantum systems.
%

\section{Acknowledgements}
This work was supported in part by JST ERATO~(No.\ JPMJER1601) and by MEXT Q-LEAP~(No.\ JPMXS0118068682).

\appendix
\section{EXPERIMENTAL APPARATUS}
\label{app:exp}

\begin{figure}[htbp]
  \centering
  \includegraphics[width=5cm,pagebox=cropbox, clip]{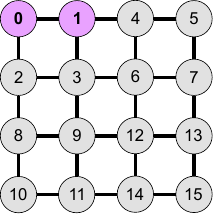}
  \caption{Schematic of the 16-qubit square lattice. The qubits used in the experiments are shaded in purple.}
  \label{fig:circuit}
\end{figure}
The device studied in this work consists of 16 qubits arranged in a 4$\times$4 square lattice. The schematic and connectivity of the device are shown in Fig.~\ref{fig:circuit}(a). The experiments are run on two of the qubits, Q0 and Q1. Individual qubit parameters are denoted in Table~\ref{tab:quantum_system_parameters}. We treat Q0 as the target qubit and Q1 as the spectator qubit. During the experiments, other qubits were not controlled or measured. The qubits are dispersively read out through coplaner waveguide resonators with Purcell filters~\cite{Evan2014}, and the readout is frequency-multiplexed for four qubits in one unit. We utilize an impedance-matched Josephson parametric amplifier~(ImPA)~\cite{Mutus2014, Tanay2015} for the qubit readout. 

\begin{table*}[htbp]
\centering
\caption{Summary of qubit parameters.}
\label{tab:quantum_system_parameters}
\begin{tabular}{c|wc{2cm}wc{2cm}}
\hline\hline
 & {Q0} & {Q1} \\ % Column swap
\hline
{Qubit frequency, $\omega_{\mathrm{q}}/2\pi$ (GHz)} & 8.508 & 7.943 \\ % Column swap
{Qubit anharmonicity, $\alpha/2\pi$ (MHz)} & $-432$ & $-357$ \\ % Column swap
% {Qubit-Qubit coupling} & \multicolumn{2}{c}{$-0.73$} \\
{Qubit energy relaxation time, $T_1$ ($\mu$s)} & 13.2 & 25.3 \\ % Column swap
% {$T_2^{\mathrm{ramsey}}$ ($\mu$s)} & 5.4 & 13.8 \\ % Column swap
{Qubit echo dephasing time, $T_2$  ($\mu$s)} & 10.4 & 15.9 \\ % Column swap
{ZZ interaction, $\xi_{\mathrm{ZZ}}/2\pi$ (MHz)} & \multicolumn{2}{c}{$-0.73$} \\
{Readout drive frequency, $\omega_{\mathrm{r}}/2\pi$ (GHz)} & 10.203 & 9.900 \\ % Column swap
{Resonator dispersive shift, $\chi/2\pi$ (MHz)} & $-1.4$ & $-0.8$ \\ % Column swap
{Resonator external decay rate, $\kappa_{\mathrm{ext}}/2\pi$ (MHz)} & 6.3 & 4.5 \\ % Column swap
{Readout error rate (\%)} & 3.8 & 3.3 \\ % Column swap
{Thermal excitation rate (\%)} & 5.5 & 8.7 \\ % Column swap
\hline\hline
\end{tabular}
\end{table*}

\section{CALIBRATION}
\label{app:cal}
\begin{figure*}[htbp]
  \centering
  \includegraphics[width=17cm,pagebox=cropbox, clip]{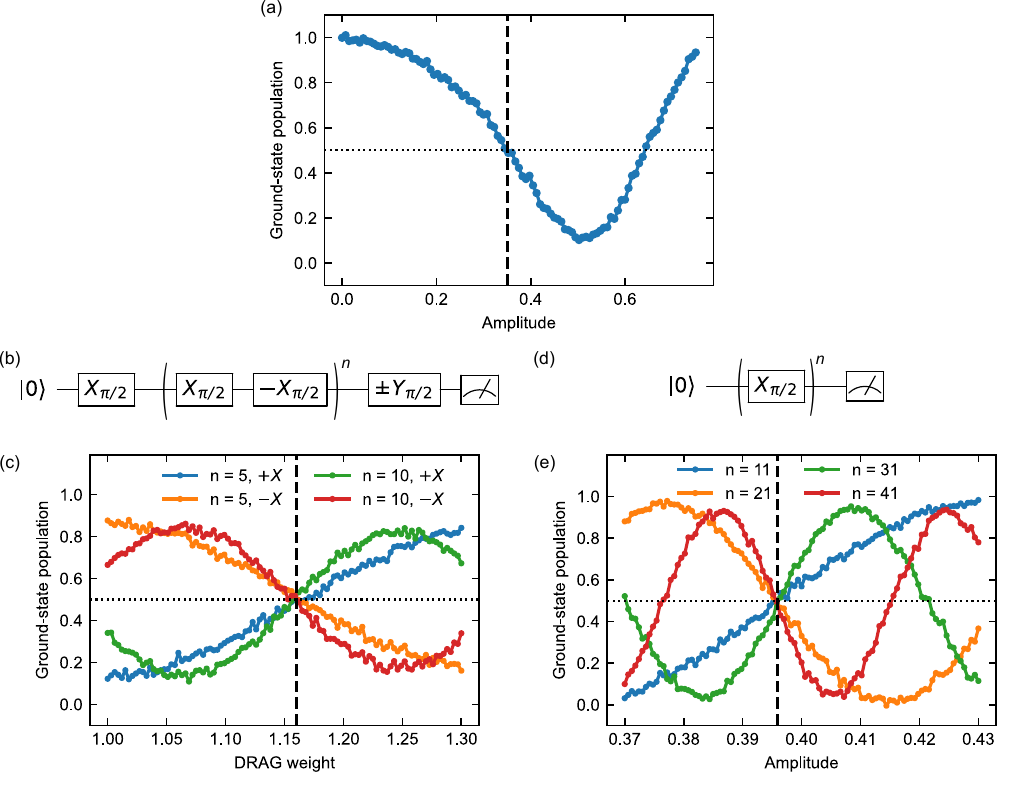}
  \caption{Experimental calibration of the designed $X_{{\pi}/{2}}$ gate on Q0. (a) Rough estimation of the pulse amplitude. (b) Gate schematic and (c) the results of DRAG-weight calibration. (d) Gate schematic and (e) the results of pulse-amplitude calibration. The vertical dashed lines and the horizontal dotted lines in~(a),~(c) and~(e) indicate the calibrated values and the ground-state population $P_{\mathrm{g}}=0.5$, respectively.}
  \label{fig:cal}
\end{figure*}

We apply the $Y$-only DRAG presented in Ref.~\citenum{Gambetta2011} to suppress unwanted effects due to the presence of non-computational states. We first roughly estimate the pulse amplitude for $X_{{\pi}/{2}}$ gate as shown in Fig.~\ref{fig:cal}(a). We select an amplitude that results in a ground-state population closest to 0.5 after the pulse is applied. Although the amplitude-population curve should be cosine-like, this is not the case here. We suppose this is because when strongly driven, the unwanted interaction is induced between the target and surrounding qubits. We repeat the DRAG-weight and pulse-amplitude calibrations alternately until the optimal condition remains constant. In Fig.~\ref{fig:cal}(b), we show the gate schematic for calibrating the DRAG weight, which is based on Ref.~\citenum{Lucero2010}. In the pulse sequence, we apply $(X_{{\pi}/{2}}\ X_{-{\pi}/{2}})^n$ to the initial state $(\ket{0}-i\ket{1})/\sqrt{2}$ so that the phase error is amplified, and then measure $\braket{\pm\sigma_x}$. We run this pulse sequence with sweeping the DRAG weight to determine the optimal point where $\braket{\pm\sigma_x}=0$ for any $n$. Examples of the DRAG-weight calibration results are shown in Fig.~\ref{fig:cal}(c). Fig.~\ref{fig:cal}(d) shows the pulse sequence to finely calibrate the pulse amplitude. In this sequence, we apply $(X_{{\pi}/{2}})^n$ to the initial state $\ket{0}$ and measure the ground-state population. The optimal pulse amplitude is the point where the ground-state population is 0.5 for any odd $n$. 

\section{COMPARISON WITH COMPOSITE PULSE}
\label{app:cp}
\begin{figure}[H]
    \centering
    \includegraphics[width=8.6cm,pagebox=cropbox, clip]{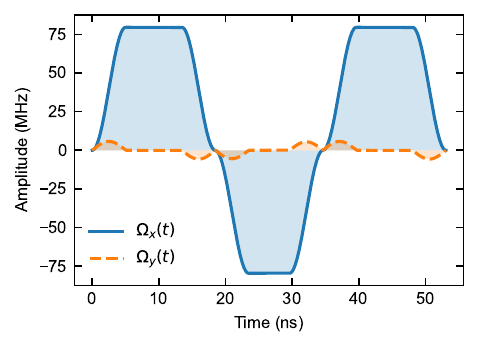}
    \caption{Quadrature components of the waveform of the CORPSE $X_{\pi/2}$ gate with DRAG correction.}
    \label{fig:corpse_waveform}
\end{figure}

\begin{figure}[H]
    \centering
    \includegraphics[width=8.6cm,pagebox=cropbox, clip]{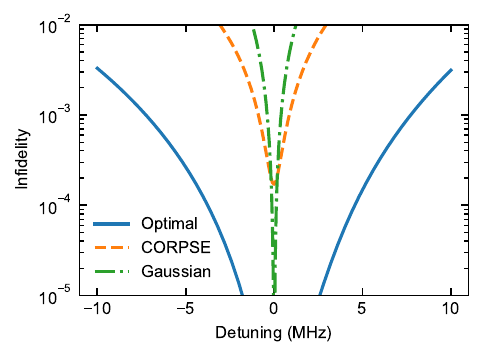}
    \caption{Comparison of the $X_{\pi/2}$ gate in the robustness against the drive detuning with the CORPSE, optimal, and Gaussian pulses.}
    \label{fig:corpse_result}
\end{figure}

Composite pulses are known as a way to achieve quantum gates robust against systematic errors. Here, we compare the performance between our optimal gate and a composite pulse family called CORPSE~(Compensation for Off-Resonance with a Pulse SEquence)~\cite{Cummins2003} that is robust against detuning error. In general, CORPSE assumes that the individual pulses are rectangular shapes of equal amplitude. However, rectangular pulses are hard to execute because of the bandwidth limitation and the discontinuous turning-on and off of the pulses. Here, we instead utilize flat-top raised-cosine pulses to relax the sharp edges. This modification results in the deviation of the time evolution from the original theory and the reduction of the robustness. The rotation angles of the three pulses that construct the CORPSE $X_{\theta}$ gate are given by~\cite{Cummins2003}
\begin{align}
    \theta_1 &= 2n_1\pi+\frac{\theta}{2}-\arcsin{\frac{\sin{(\theta/2)}}{2}}\,,\\
    \theta_2 &= 2n_2\pi-2\arcsin{\frac{\sin{(\theta/2)}}{2}}\,,\\
    \theta_3 &= 2n_3\pi+\frac{\theta}{2}-\arcsin{\frac{\sin{(\theta/2)}}{2}}\,,
\end{align}
where $n_1$, $n_2$ and $n_3$ are integers, subject to the physical restriction that the resulting pulse angles must be positive. While choosing $n_1=1$, $n_2=1$ and $n_3=0$ is best since it minimizes remaining second-order error~\cite{Cummins2003}, such a choice makes the third pulse too short and experimentally impractical in terms of the adiabaticity. We thus choose $n_1=n_2=n_3=1$, which is still first-order insensitive against detuning error. The rotation angles of the three pulses that construct the CORPSE $X_{\pi/2}$ gate are given by \{$2.135\pi, 1.770\pi, 2.135\pi$\}, and pulse phases are \{$0, \pi, 0$\}, respectively. We set the rise-up and ring-down times to be 5~ns each. The maximum amplitudes of each pulses are the same as the optimal pulse used in the experiment. Figure~\ref{fig:corpse_waveform} shows the waveforms of the CORPSE composite pulses. Each pulse includes numerically-optimized DRAG correction. The comparison in robustness against the detuning error is shown in Fig.~\ref{fig:corpse_result}. Although the CORPSE composite pulse shows better robustness against detuning than the Gaussian pulse, the gate fidelity at zero detuning is orders of magnitude worse than the optimal and Gaussian pulses. This inferiority is due to the significant leakage error caused by the sharp edges and broad spectral distribution. If the rise-up and ring-down times of the flat-top pulses are made slower to reduce the leakage error, the advantage of the composite pulse is lost because the robustness is reduced. Therefore, the comparison made in this section highlights the advantage of our proposal over the composite pulse. Our method combines practicality in experimental applications with robustness against errors thanks to the careful choice of the waveform ansatz.

\end{document}